# An Alternative Explanation of Supernova Ia Data

Yijia Zheng

National Astronomical Observatory, Chinese Academy of Sciences, Beijing, 100012 China

## Abstract

Before 1998 the universe expansion was thought to be slowing down. After 1998 the universe expansion is thought to be accelerating up. This change of the belief is motivated by the observed brightness of the high redshift supernova Ia fainter than expected. In this paper it is argued that this change of the belief is not necessary. There is a more reasonable explanation for the observed brightness of the supernovae Ia fainter than expected. That is: the universe space filled with the tenuous ionized gas; the Thomson Effect of free electron in the ionized gas caused the observed brightness of the high redshift supernova Ia fainter than expected. The observations of Warm-Hot Intergalactic Medium (WHIM) support this explanation. The universe space filled with tenuous ionized gas explanation will lead to the puzzle of the dark matter and dark energy disappeared simultaneously.

## 1. INTRODUCTION

Before 1998 the universe expansion was thought to be slowing down. After 1998 the universe expansion is thought to be accelerating up. This change of the belief is motivated by the observed brightness of the high redshift supernovae Ia fainter than expected [1, 2]. The discovery of the observed brightness of the high redshift supernovae Ia fainter than expected is of very great importance to understand the universe. But the explanation of the accelerating universe is questionable. This explanation means that the universe space is filled with the baffling dark energy; the dark energy must have negative pressure $p$. Einstein in 1917 thought that the negative pressure $p$ is an unphysical result and tried to avoid this unphysical result by introducing a cosmological constant [3, 4]. The big bang model of the universe was based on the assumption that the cosmological constant is zero. Today, to explain the accelerating expansion of the universe, astronomers think that the cosmological constant must exist and it must have a positive value; it represents the baffling dark energy.

Meanwhile this explanation of the accelerating universe needs to suppose that the universe space is completely transparent for the light of high redshift supernova Ia. This is a loophole. Perlmutter 1999 had noticed this loophole and attempted to plug in it [4]. However he thought that only the dust in the universe space can dim the light of the high redshift supernova Ia. His conclusion was: 'Fortunately, there is a straightforward way to measure this. All dust that has so far been observed in the universe absorbs blue light more than red light, so sources that are seen through dust appear redder. We can compare the colors of the nearby supernovae and the distant supernovae to see if the latter are in fact redder. We find no statistical difference, and a full statistical analysis indicates that ordinary dust that reddens cannot account for our supernova data without a cosmological constant'.

Weinberg in his book 'gravitation and cosmology' (15.2) proposed: 'if one tentative accepts the result that the $q_o$ is of the order unity, then one is forced to the conclusion that the mass density of about $2 \times 10^{-29}$ g/cm$^3$ must be found somewhere outside the normal galaxies'. After considered several possibilities, Weinberg wrote: 'it is more conservative to suppose that the missing mass takes the form of a tenuous hydrogen gas, ionized or neutral, filling all space' [3]. In his book (15.4), using the Penzias and Wilson repot and adopt $H_0$=75 km/sec/Mpc, Weinberg estimated that the neutral hydrogen gas number density is less than $3 \times 10^{-6}$ cm$^{-3}$. If the hydrogen gas is ionized, Weinberg wrote: 'For example, suppose that $q_o$=1 and $H_0$=75 km/sec/Mpc. We then expect at present electron number density $n_{eo}$ = $1.2 \times 10^{-5}$ cm$^{-3}$.'

It is hard to determine the mass density in the universe space by astronomical observations. Different astronomers give the mass density of the universe enormous differences. For example, Loh and Spillar 1986 estimate that the density parameter $\Omega$ = 0.9, consistent the Einstein-de Sitter model [5]. In 1998 Fukugita et al. present an estimate of the global budget of baryons in all states. They concluded that most of the baryons today are still in the form of ionized gas [6]. Fang et al., 2010 confirmed that the total amount of the luminous baryons in the nearby universe accounts for at most 50% of the total baryonic matter in the low-redshift universe [7]. But their total budget of baryons, expressed as a fraction of the critical Einstein-de Sitter density, is in the range

$0.007 \leq \Omega_B \leq 0.041$. The estimation of the mass density of the universe strongly depends to the assumption of the universe model.

The total budget of baryons estimated by Fukugita et al. evidently is underestimated. They did not consider the contribution of baryons in the intercluster space. For intercluster space, the electron density is lower than that in the clusters and hard to be detected. But this does not mean that their contribution to the total budget of baryons in the universe is small. In rich clusters, the electron number density is large enough, so it is easy to be detected. For example, Planck Collaboration 2013 gave the electron number density in the center region of the Coma cluster is $2.9 \times 10^{-3}$ cm$^{-3}$ [8]. But these baryons' contribution to the total budget of baryons is small, because the volume of rich clusters only occupies a small part of the universe. Weinberg in his book (15.2) pointed out: 'If the missing mass is not within clusters of galaxies, then we must look for it in the space between the clusters. One reasonable requirement is that the total density of intercluster must be less than the density within cluster, so that the clusters represent appreciable condensations. The total volume outside clusters is roughly 500 times (15.2.13). Hence, that even if the density outside clusters is an order of magnitude less than the density within cluster, there is still plenty of room in intercluster space for all the missing mass we need'. Recent numerical simulations and astronomical observations on the Warm-Hot Intergalactic Medium (WHIM) support Weinberg's view point.

Numerical simulations have shown that a large fraction of baryonic material in the universe is contained within the Warm-Hot Intergalactic Medium in large-scale, filamentary structures comprised of diffuse gas [9, 10]. By using the soft X-ray emission Fraser-McKelvie et al. (2011) derived an averaged electron density for filaments of $N_e=(4.7\pm0.2)\times10^{-4} h_{100}^{1/2}$ cm$^{-3}$ in the 0.9–1.3 keV band, or in the 0.5–2.0 keV band $N_e=(9.6\pm0.2)\times10^{-4} h_{100}^{1/2}$ cm$^{-3}$ [11]. Significantly, Aragon-Calvo et al. (2010) suggest that filaments occupy 10 per cent of the Universe by volume [11, 12]. In this case the density parameter $\Omega$ of the baryon may be of the order of unity.

Various techniques can be used to detect the missing baryons but should be with improvements in sensitivity [9]. Actually, using the Thomson scattering effect of the free electron to observe the light dimming of distant astronomical objects is a very effective method to detect the electron number density in the universe space. But most astronomers have a wrong impression that the Thomson scattering effect of the free electron can be neglected. They believe that the total baryon budget expressed in the density parameter Ω is only about 0.04. In this case, the observed brightness of high redshift supernovae Ia is impossible caused by the Thomson scattering effect of the free electron in the universe.

If the average electron number density in the universe space is of the order $N_e = 1.2 \times 10^{-5} cm^{-3}$, then 'the observed brightness of high redshift supernovae Ia is fainter than expected' can be explained as: the observed light from the high redshift supernovae Ia is dimmed by the tenuous ionized gas in the universe space due to the Thomson scattering effect of the free electron. The observed results of Fraser-McKelvie et al. (2011) imply that the average electron number density in the universe space is possible of this magnitude order [11, 12]. This average electron number density in the universe space is consistent with Weinberg's estimate on the missing mass. This explanation leads to the puzzle of dark energy and dark matter disappeared simultaneously. In the next section the calculated dimmed brightness of the high redshift supernovae Ia caused by Thomson scattering effect of the free electron is shown in the Hubble diagram. The result is equivalent to the expansion of the universe is accelerating [1].

## 2. HUBBLE DIAGRAM

The astronomer's Hubble diagram is a plot of the observed magnitude against redshift. The Hubble diagram is interpreted as follows: for an object of known brightness, the fainter the object the farther away it is and the further back in time you are looking, so you can treat the y-axis as the time axis. The x-axis, the redshift, is a very direct measurement of the relative expansion of the universe, because as the universe expands the wavelengths of the photons traveling to us stretch exactly proportionately—and that is the redshift. Thus the Hubble diagram is showing you the "stretching" of the universe as a function of time. As you look farther and farther away, and further back in time, you can find the deviations in the expansion rate. Therefore, the brightness of high redshift supernova Ia fainter than expected leads Perlmutter to conclude that the universe expansion is accelerating up [4].

For a flat universe, if the universe is complete transparent, the relation between the observed magnitude $m$ and the absolute magnitude $M$ is

$$m = M - 5\log\frac{R_0}{R} \qquad (1)$$

where $R_0 = 10$ pc, $R$ the luminosity distance of the observed object.

If in the universe space is filled with the ionized gas, the formula (1) should be modified. The observed light of the object will be dimmed by electrons in the ionized gas due to the Thomson scattering. Suppose the average electron number density in the universe is $N_e$, then the optical depth $\tau$ of the observed object (caused by Thomson scattering) is

$$\tau = N_e \sigma_c R \qquad (2)$$

where $\sigma_c$ is the Thomson scattering cross section. The formula (1) should be modified to

$$m = M - 5\log\frac{R_0}{R} + 2.5\frac{\tau}{\ln(10)} \qquad (3)$$

The affect of the ionized gas in the universe space lead to observed magnitude of the object increase (fainter) by

$$\Delta m = 2.5\frac{\tau}{\ln(10)} \qquad (4)$$

This is why the observed brightness of the high redshift supernova Ia fainter than expected. This affect caused by ionized gas in the universe space did not considered by Perlmutter [4].

The optical depth of the observed object is proportional to the luminosity distance of the observed object; according to Hubble law it is proportional to the redshift of the observed object too. In the nearby universe, the redshift of the objects are small, so the affect of the ionized gas is small and it is very hard to be detected. For high redshift supernova Ia, due to the optical depth increasing with the distance, the affect of the ionized gas in the universe space becomes large, so it is very easy to be detected.

Because the optical depth of an observed object is proportional to the redshift of the observed object, the optical depth $\tau$ can be connected to the redshift $Z$ by

$$\tau = \alpha \cdot Z \qquad (5)$$

where α is a ratio constant. Figure 1 is a Hubble diagram which showed the observed brightness variation caused by the ionized gas. The black line correspond to the ratio constant $a=0$. This means the universe space is complete transparent. The universe expansion obeys the Hubble law. The green line (α=0.5) and the red line (α=1) show the effect of the Thomson scattering caused by the ionized gas in the universe space. The result is equivalent to the affect caused by the accelerating expansion of the universe [1].

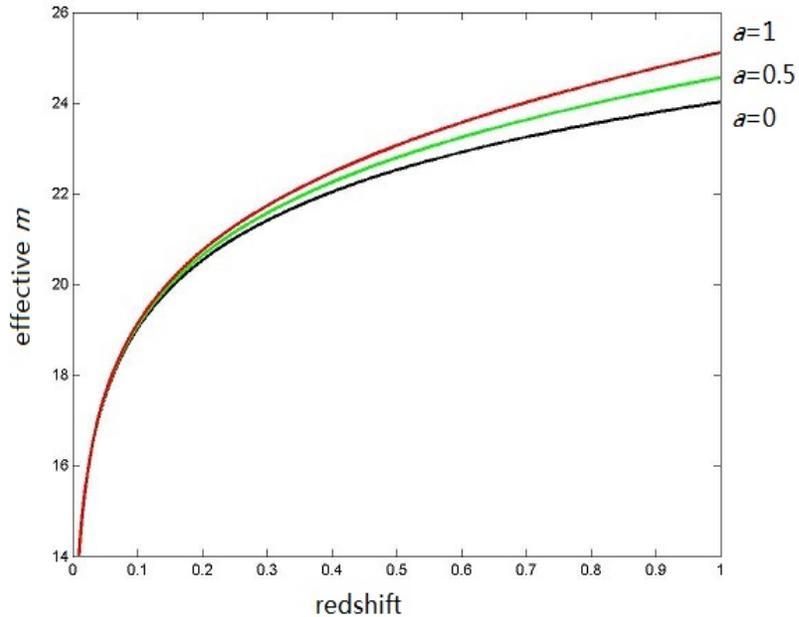

Figure 1, The Hubble diagram showed the observed brightness variation caused by the ionized gas

Because $\alpha = N_e \sigma_c / H_0 / C$, Suppose $H_0$=75 km/s/Mpc, then $\alpha$=1 means $N_e$=1.3x10$^{-5}$ cm$^{-3}$. It is consistent with the critical Einstein-de Sitter mass density.

## 3. CONCLUSION

At present, most astronomers believe that the observed brightness of high redshift supernovae Ia fainter than expected is caused by the universe accelerating expansion. This explanation requires that universe space must fill with the dark energy; the dark energy must have negative pressure $p$. Einstein in 1917 thought that the negative pressure $p$ is an unphysical result [3, 4]. That is why the dark energy still remain as a puzzle for physics.

There is a more simple and reasonable procedure to explain the observed brightness of high redshift supernovae fainter than expected. It is that the brightness dimming is caused by the Thomson Scattering of free electrons. This explanation requires that the universe space fill with the tenuous ionized gas; the ionized gas mass density is about $2 \times 10^{-29}$ g/cm$^3$. This average ionized gas mass density in the universe space is consistent with Weinberg's estimate on the missing mass. Fraser-McKelvie et al. (2011) by using the soft X-ray emission of filaments derived the averaged electron density implies that this ionized gas mass density is a reasonable estimate [11, 12]. This explanation leading to the puzzle on the dark matter and dark energy disappear simultaneously. Hence, This explanation is simple and more reasonable.

## Acknowledgment

I would like to thank Dr. Nailong Wu for the correction and suggestions his made to greatly improve the English of the manuscript.